\documentclass[12pt,preprint]{aastex}
\lefthead{Skinner et al.}
\righthead{NGC 2071}

\begin{document}

\title{Hard X-rays and  Fluorescent Iron Emission from the 
       Embedded Infrared Cluster in NGC 2071}

\author{Stephen L. Skinner, Audrey E. Simmons}
\affil{CASA, Univ. of Colorado, Boulder, CO, USA 80309-0389 }

\author{Marc Audard}
\affil{Integral Science Data Centre, Ch. d'Ecogia 16, CH-1290 Versoix,
Switzerland}
\affil{Geneva Observatory, University of Geneva, Ch. des Maillettes 51,
       1290 Sauverny, Switzerland}

\author{Manuel  G\"{u}del}
\affil{Paul Scherrer Institute, W\"{u}renlingen and Villigen, 
       CH-5232 Switzerland}

%
\newcommand{\ltsimeq}{\raisebox{-0.6ex}{$\,\stackrel{\raisebox{-.2ex}%
{$\textstyle<$}}{\sim}\,$}}
%
\newcommand{\gtsimeq}{\raisebox{-0.6ex}{$\,\stackrel{\raisebox{-.2ex}%
{$\textstyle>$}}{\sim}\,$}}

\begin{abstract}
We present first results of {\em XMM-Newton}  X-ray observations 
of the  infrared cluster lying near the NGC 2071 reflection nebula 
in the Orion B region. This cluster is of interest  because it is 
one of the closest regions known to harbor embedded high-mass stars.
We report the discovery  of hard X-ray
emission from the dense central NGC 2071-IR subgroup which contains at least
three high-mass young stellar objects (NGC 2071 IRS-1, IRS-2, and IRS-3).
A prominent X-ray source is detected  within 1$''$ of the  infrared
source IRS-1, which is thought to drive a powerful bipolar molecular
outflow. The X-ray spectrum of this source  is quite unusual compared 
to the optically thin plasma spectra normally observed in young stellar
objects (YSOs). The spectrum is characterized by a hard broad-band 
continuum plus an exceptionally broad emission line at  $\approx$6.4 keV 
from  neutral or near-neutral iron. The fluorescent Fe  line
likely originates in cold material near the embedded star (i.e. a
disk or envelope) that is irradiated by the hard heavily-absorbed 
X-ray source.
\end{abstract}



%
\newpage

\section{Introduction}
A high-mass star can arrive on the main-sequence completely 
enshrouded in dust and inaccessible to optical studies. Thus,
the ability to penetrate high extinction is crucial to exploring
the earliest stages of high-mass star formation. Infrared, radio, 
and millimeter observations have traditionally been used and 
provide important information on physical conditions in circumstellar
disks and envelopes  or in ionized winds or HII regions around the massive
young star. X-ray observations can also penetrate high extinction and
provide a  different perspective that probes high-energy 
processes including magnetic activity originating close to the 
stellar surface, mass-loss as traced by  shocked winds,
jets, or outflows, and hot diffuse gas that   
pervades some young clusters containing massive young OB
stars with powerful winds.

One of the closest regions known to contain young high-mass stars
is the infrared cluster near the optical reflection 
nebula NGC 2071 in the Orion B region
(Lynds 1630 dark cloud) at a distance of $\sim$400 pc (Anthony-Twarog 1982;
Brown et al. 1994). Near-infrared observations by Lada et al. (1991)
revealed more than 100 K-band sources in a 100 arc-min$^{2}$ region 
down to  K $\approx$ 14 mag. This extended cluster surrounds a dense
central subgroup  known as NGC 2071-IR which contains at least 10 
near-IR sources in a $\approx$1$'$ $\times$ 1$'$ region 
(Walther et al. 1993 = Wa93). Of particular interest in the 
NGC 2071-IR subgroup are IRS-1, IRS-2,  and IRS-3. These sources 
are surrounded by compact HII regions (Snell \& Bally 1986)
and H$_{2}$O and OH masers lie in proximity. More detailed
information on NGC 2071-IR and properties of the infrared sources
can be found in  Aspin, Sandell, \& Walther (1992) 
and Wa93, and references therein. The presence of strong ionizing
UV radiation and maser emission strongly suggests that these objects
are young embedded early-type stars. 
Such objects are exceedingly rare at distances less than 500 pc
and the NGC 2071 cluster thus plays a key role in observational 
studies of high-mass star formation.

We report here on first results of a pointed X-ray observation
of NGC 2071 obtained with {\em XMM-Newton}. This observation is
centered on the  NGC 2071-IR subgroup and provides broader
energy coverage and improved spectral information compared to two
previous {\rm ROSAT} HRI exposures  (rh202521/22) 
which captured the IR subgroup $\approx$8$'$ - 10$'$ 
off-axis. Our objectives were to (i) obtain an X-ray census of 
the NGC 2071 cluster and (ii) search for X-ray emission from 
the massive embedded stars, whose X-ray properties are largely
unknown. We report the detection of hard emission lying within 1$''$ of 
NGC 2071-IRS1 and discuss its very unusual X-ray spectrum
which is dominated by strong fluorescent Fe line emission.

\section{XMM-Newton Observations}
The {\em XMM-Newton} observation  began on
2005 March 30 at 15:20 UT and ended on March 31 at 03:53 UT.
Pointing was centered on NGC 2071-IRS 1 (Wa93).  
The  European Photon Imaging Camera (EPIC) provided 
CCD imaging spectroscopy from the pn camera 
(Str\"{u}der et al. 2001) and two nearly
identical MOS cameras (MOS1 and MOS2;
Turner et al. 2001). The medium optical blocking
filter was used. The EPIC cameras provide
energy coverage over E $\approx$ 0.2 - 15 keV with 
energy  resolution E/$\Delta$E $\approx$ 20 - 50.
The MOS cameras provide the best on-axis angular 
resolution with FWHM $\approx$ 4.3$''$ at
1.5 keV.

Data were reduced using the {\em XMM-Newton}
Science Analysis System (SAS vers. 6.1). Event
files generated by  {\em XMM-Newton} standard
processing  were time-filtered to remove the first
$\approx$15 ksec of data, which were affected by
high background radiation.  
This yielded  29.9 ksec of usable pn exposure
and 30.1 ksec of usable exposure per MOS.
Source detection was accomplished with the
SAS task {\em edetect\_chain} on images filtered
in different bands. These results were compared
with the source list provided from the {\em XMM-Newton}
pipeline processing and images were visually checked
for missed or spurious detections. The  properties
of sources detected in the central cluster region
(Table 1) are based on events in the 0.5 - 7.5 keV range.

Spectra and light curves were extracted from  circular
regions of radius  R$_{e}$ = 15$''$ centered
on individual sources,  corresponding to $\approx$68\%
encircled energy at 1.5 keV. Smaller regions were used
for a few closely spaced sources to avoid region overlap.
Background files were
extracted from circular source-free regions near the source.
The SAS tasks {\em rmfgen} and {\em arfgen} were used to
generate source-specific response matrix
files (RMFs) and auxiliary response files
(ARFs) for spectral analysis. The data were
analyzed using the {\em XANADU} software
package
\footnote{The {\em XANADU} X-ray analysis software package
is developed and maintained by NASA's High Energy
Astrophysics Science Archive Research Center (HEASARC). See
http://heasarc.gsfc.nasa.gov/docs/xanadu/xanadu.html
for further information.},
including {\em XSPEC} vers. 12.3.0.


\section{X-ray Overview of NGC 2071}
Figure 1 shows a broad-band 0.5 - 7 keV EPIC pn image of
the central $\approx$10$'$ $\times$ 10$'$ cluster region surrounding
the NGC 2071-IR subgroup. Analysis of both the pn and 
MOS images resulted in 33 X-ray detections in this central 
region, as summarized in Table 1. 
Possible counterparts were
found for 23 of these 33 sources (70\%) within a search 
radius of 2$''$ using the 2MASS, HST GSC v2.2, and
USNO B1 electronic data bases. Of the 23 sources in Table 1
with identifications, 22 have 2MASS counterparts and their K$_{s}$ magnitudes
are in the range K$_{s}$ = 6.3 - 14.1. The X-ray positions in 
Table 1 have been registered against 2MASS and the mean positional
offset between the  X-ray sources and their  assigned counterparts 
is only  0.8$''$. A view of the  harder X-ray sources in the cluster
is seen in the 4 - 7 keV  image (Figure 2). The most prominent
hard-band detections are a  source lying within 
1$''$ of  IRS-1 (source 13 = XMM J054704.78$+$002143; Figs. 2 and 3),
a source associated with  2MASS J054705.25$+$002253
(source 16), a close X-ray pair lying
near the position of IRAS 05445$+$0016 (sources 20 and 21; Fig. 4),
and variable X-rays (Fig. 5) from the emission-line star
LkH$\alpha$ 308 (source 19). 
These hard sources are discussed further below.

Sufficient counts were present in 21 sources to obtain fits
of  the X-ray spectra with either one-temperature (1T) or
two-temperature (2T) $apec$ optically thin plasma models. The 
absorbed fluxes for these objects based on spectral fits
are given in Table 1 and fluxes for fainter sources were
estimated using PIMMS\footnote{Further information on the
Portable Interactive Multi-Mission Simulator (PIMMS)
can be found at 
http://heasarc.gsfc.nasa.gov/docs/software/tools/pimms.html.}.
Two-temperature models gave acceptable fits of 10 of the brightest
sources, excluding source 13 which is discussed in detail
below (Sec. 4). These 10 sources (source numbers 
1,4,8,12,14,18,19,20,21,29)  gave a median hydrogen column
density  log N$_{\rm H}$ = 22.0 cm$^{-2}$ and median plasma
temperatures  kT$_{1}$ = 0.74 keV and kT$_{2}$ = 2.8 keV.
The means are nearly identical to the medians.

\subsection{The NGC 2071-IR Region}
Figure 3 shows the summed  MOS1$+$2 image of the 
NGC 2071-IR region known to contain embedded massive young stars. An
unusual X-ray source  (source 13) is 
nearly coincident with IRS-1. Its  offset from
the VLA position of IRS-1 (Torrelles et al. 1998) and 
the near-IR source 2MASS J054704.78$+$002142 is only 0.$''$7. 
Thus, an association of this X-ray source with IRS-1 is likely.
A search of the HEASARC galaxies data base
\footnote{http://heasarc.gsfc.nasa.gov/cgi-bin/W3Browse/w3browse.pl}
revealed no known galaxies or AGNs within 20$''$ of the IRS-1 position 
so the probability of a chance association between this X-ray source and
a distant background  object is small. A catalogued VLA 20 cm radio source 
NVSS J054705.01$+$002147.2 lies 5.$''$1 northeast of the X-ray 
peak and is most probably associated with IRS-5, which is offset by
only 0.$''$7 from the VLA source.

The X-ray source located within 1$''$ of IRS-1   
has 262 net pn counts (0.5 - 7.5 keV; R$_{e}$ = 15$''$) and its mean
photon energy $<$E$>$ = 4.62 keV is the highest of any source
in  Table 1. Not only is it visible in the 
4 - 7 keV hard-band image (Fig. 2), it is also seen in images
restricted to the higher 8 - 12 keV range. No large amplitude
flare-like variability is seen in the X-ray light curve of this source
(Fig. 6) but fluctuations at the $\approx$2$\sigma$ level are
present. A  $\chi^2$ test using the pn light curve gives a probability 
of constant count rate
rate  P$_{const}$ = 0.13 ($\chi^2$/dof = 19.8/14; bin size = 2000 s),
so no significant variability can be claimed.
Fainter X-ray emission is visible in Figure 3 extending to the
northeast of IRS-1 that may be associated with IRS-2
($\delta$ = 11.$''$4), IRS-3 ($\delta$ = 5.$''$6),
or IRS-5 ($\delta$ = 6.$''$2) where the offsets $\delta$
are relative to IRS-1. In addition,
faint emission is present near the position of HH 437 (Zhao et al. 1999)
which lies $\approx$15$''$ northeast of IRS-1. 
Higher angular resolution
images will be needed to make unambiguous identifications 
for this fainter emission.

\subsection{The IRAS 05445$+$0016 Region}
The luminous far-IR source IRAS 05445$+$0016 is located
$\approx$ 4$'$ south of IRS-1, but with rather large
IRAS position uncertainties (Fig. 4). This region is of 
interest because of reported maser detections in the vicinity
(e.g. source Onsala 59 in Harju et al. 1998).
The IRAS source was listed as a candidate pre-main sequence
object by Clark (1991). Additional far-IR scans of NGC 2071
at 50 $\mu$m and 100 $\mu$m  were obtained with the 
{\em Kuiper Airborne Observatory} by Butner et al. (1990).

A close pair of  2MASS sources separated
by $\approx$15$''$ lies on either side of the IRAS 
position (Fig. 4) and both were detected by {\em XMM-Newton}
(sources 20 and 21). Both sources are visible in 
the hard-band image (Fig. 2). The southern source
(source 20 = XMM J054707.66$+$001740) has the highest
mean count rate of all sources in Table 1. Its
X-ray emission is clearly variable as discerned by
a slow decay in the pn light curve during the observation.
Its pn spectrum  shows a hard component including
emission from the Fe K complex ($\approx$6.7 keV). We were able
to obtain a good spectral fit with an absorbed two-temperature
optically thin {\em apec} plasma model with an 
absorption column density log N$_{\rm H}$ = 21.8 cm$^{-2}$, 
plasma temperature components at kT$_{1}$ = 0.87 keV
and kT$_{2}$ = 3.1 keV, and $\chi^2$/dof = 94.8/97.  The combination
of X-ray variability and a high-temperature component 
(T$_{2}$ $\approx$ 36 MK) are
characteristic of magnetic  activity. The
northerly source (source 21 = XMM J054707.93$+$001755)
is offset by only 0.$''$8 from the B-type star HDE 290861 
(V1380 Ori), a known eclipsing binary system with a
component separation of 0.$''$59 at position angle (PA)
$\approx$217$^{\circ}$ (Prieur et al. 2001).
Higher spatial  resolution observations will be needed
to determine whether the X-rays come from the B star 
itself or the  companion.

\subsection{LkH$\alpha$ 308}
The second brightest X-ray detection  in Table 1 is
LkH$\alpha$ 308 (source 19 = XMM J054707.29$+$001932).
This V = 15.6 mag emission-line star was identified  
as a probable T Tauri star by Herbig \& Kuhi (1963)
and H$\alpha$ emission was confirmed by 
Wiramihardja et al. (1989). It was discovered to be a 
relatively bright infrared source by Strom, Strom, \&
Vrba (1976) and 2MASS data give K$_{s}$ = 8.3 mag.
LkH$\alpha$ 308 is an X-ray variable as discerned from
its pn light curve which shows a slow rise and fall
in count rate by a factor of $\sim$2 during 
the observation (Fig. 5). The pn spectrum
reveals a faint Fe K emission line and spectral
fits with an absorbed two-temperature optically
thin plasma model require  a hot component at
kT$_{2}$ $\approx$ 3.5 - 5 keV as commonly seen in
magnetically active T Tauri stars.

\section{X-ray Spectrum of IRS-1}
The X-ray spectrum of the source that we associate 
with NGC 2071-IRS1  is unusual
compared to the  optically thin plasma spectra typically seen
in young stellar objects. As Figure 7 shows, the spectrum is characterized
by a nearly flat broad-band continuum that is heavily absorbed below 
$\approx$1 keV and a strong broad emission line from neutral or 
near-neutral iron near 6.4 keV. The fluorescent Fe line 
dominates the spectrum and its intensity and width are 
exceptional for a young stellar object. 

The spectrum in Figure 7 was extracted using a circular region
of radius R$_{e}$ = 15$''$ and may include contributions from 
the nearby sources  IRS-2, IRS-3, and IRS-5, which lie at offsets 
of 5.$''$6 - 11.$''$4 from IRS-1. However, we were able to recover
a nearly identical spectrum including the strong fluorescent Fe line 
when using a smaller extraction region of radius R$_{e}$ = 4$''$.
The positions of IRS-2, IRS-3, and IRS-5 lie outside this smaller 
circle and even though there could be some PSF overlap these
results suggest that IRS-1 (or an as yet unresolved source within
a few arcseconds  of IRS-1) is the dominant X-ray contributor.

We attempted to fit the spectrum using a conventional optically
thin plasma model with a single absorption component plus a Gaussian 
line near 6.4 keV. This model ran away to unphysically high temperatures 
even when multiple temperature components were allowed. However, we were
able to fit the spectrum with either (i) an absorbed power-law continuum with a
photon power-law index $\alpha_{ph}$ = $+$0.55 plus
a Gaussian line centered at E$_{line}$ = 6.48 keV ($\chi^2$/dof = 14.3/14) 
or (ii) a two-component optically thin plasma model with a
cool moderately absorbed component and a hot heavily-absorbed
component plus a  Gaussian line centered at 
E$_{line}$ = 6.43 keV ($\chi^2$/dof = 10.7/11). 
Fit results are summarized in Table 2.

Both the power-law and thermal fits are 
formally acceptable but the thermal model
provides a better fit of the shape of the spectrum at 
lower energies between 1 - 2 keV and is easier to
justify on physical grounds.
Also, the X-ray absorption log N$_{\rm H}$ = 
22.0 cm$^{-2}$ determined from the power-law model 
corresponds to a visual extinction 
A$_{\rm V}$ = 4.5 mag (Gorenstein 1975),
which is much less than the range A$_{\rm V}$ $\approx$
28 - 51 mag expected toward IRS-1 if it is an embedded
B0 - B5 star (Wa93). The thermal model yields higher
absorptions that are consistent with the range expected
for IRS-1.  However, we do consider the possibility of
power-law models further in Section 5.

High (but physically realisitic) X-ray temperatures are 
required by the two-component 
thermal model to reproduce the  broad-band continuum
(Table 2). The absorption inferred for the hot component 
at kT$_{2}$ = 10.8 keV is log N$_{\rm H,2}$ = 23.2 cm$^{-2}$.
This N$_{\rm H,2}$  implies an equivalent
visual extinction A$_{\rm V}$ $\approx$ 70 mag
(Gorenstein 1975).  Thus, the high-temperature
source is heavily obscured.  The absorption associated with the 
cooler component at kT$_{1}$ = 0.7 keV is 
log N$_{\rm H,1}$ = 22.6 cm$^{-2}$, or 
A$_{\rm V}$ $\approx$ 17 mag. 
The strong Fe line accounts for
$\approx$30\% of the observed (absorbed) flux in the 
0.5 - 7.5 keV range.
The Gaussian line width deduced from the
thermal model $\sigma_{line}$ = 140 eV gives FWHM = 330 eV.
The line is quite likely resolved since the pn intrinsic energy 
resolution is FWHM $\approx$ 160 eV at 6.5 keV. 

\subsection{The Fluorescent Fe Line}

The physical picture needed to explain the fluorescent Fe
emission requires the presence of neutral or near-neutral
material in  proximity to the X-ray source.
This material is irradiated by the hard  source, which
is quite likely the embedded  high-mass star. The origin
of the line broadening is clearly of interest. Velocity
broadening of a single line cannot fully account 
for the line width without
invoking unrealistically high velocities. Some of the 
broadening could be due to multiple closely-spaced Fe
lines that are not spectrally distinguishable at
the pn energy resolution. In this regard, inspection of the
unfolded spectral model shows that a fainter Fe K line may 
contribute to some of the flux near 6.7 keV but this line,
if present, is masked by the broad wings of the 6.43 keV 
fluorescent line.

If the broad fluorescent line width is a column density effect 
then the inferred column density is large. 
The line equivalent width (EW) is related to
the column density of cold fluorescent material in the 
optically thin slab approximation by 
EW $\approx$ 2.3 N$_{24}$ keV  (Kallman 1995),
where N$_{24}$ is the column density of the cold matter
in units of 10$^{24}$ cm$^{-2}$. In the present case,
our pn spectrum measurements give EW = 2.4 keV
so N$_{\rm H,cold}$ $\sim$ 10$^{24.0}$ cm$^{-2}$.
The lower signal-to-noise ratio MOS spectra give a 
somewhat smaller value EW = 1.4 keV or
N$_{\rm H,cold}$ $\sim$ 10$^{23.8}$ cm$^{-2}$.
The value determined from the pn spectrum is 
at the upper limit where Kallman's 
approximation breaks down but if the 6.43 keV feature
is a blend then the inferred value of N$_{\rm H,cold}$ is 
only an upper limit. 

The above approximation indicates that the absorption column of
the cold material is about an order of magnitude greater
than the absorption inferred for the hard thermal X-ray component 
(Table 2). The material responsible for absorption of the hard 
X-rays cannot fully account for N$_{\rm H,cold}$ (see also 
eq. [4] of Tsujimoto et al. 2005). Thus, the cold fluoresced
material may not lie directly on the line-of-sight. One 
possibility is that the fluorescent line originates in
the dense ridge of molecular gas orthogonal to the outflow 
axis which may be a rotating disk (Bally 1982; Seth, Greenhill,
\& Holder 2002).

Fluorescent Fe line analysis similar to that above has been
undertaken on YSOs in other high-mass star-forming regions. Of 
particular relevance is the {\em Chandra} study of the Sgr B2 giant
molecular cloud by Takagi, Murakami, \& Koyama (2002), who reported 
high-temperature plasma and strong 6.4 keV line emission for some 
luminous X-ray sources.  Specifically, they obtained 
kT $\approx$ 10 keV and a fluorescent Fe line equivalent width 
EW = 630 (180 - 1100; 90\% confidence) eV for Sgr B2 source 10  
(CXO J174720.2$-$282305). Takagi et al.  note that this X-ray 
source lies near an ultracompact HII region and may  correspond
to a massive YSO. The X-ray temperature reported for Sgr B2 source 10
is similar to what we infer for the hot component of NGC 2071 IRS1, 
but the fluorescent line equivalent width and derived value of
N$_{\rm H,cold}$ is $\approx$3 - 4  times larger for IRS1.

\section{Discussion}
The unusual X-ray spectrum of the source near the massive young
star IRS-1 warrants further discussion. The presence of cool and
hot plasma components apparently seen through different absorption
columns suggests that more than one source or X-ray emission 
process contributes to the spectrum. We consider possible emission
processes below.

\subsection{Similarities with Jet-Driving T Tauri Stars}
The need to invoke two thermal X-ray components at different
absorption columns to fit X-ray spectra has also recently
been seen in some  accreting {\em low-mass} pre-main
sequence stars such as the T Tauri star DG Tau A (G\"{u}del 
et al. 2005). High-resolution {\em Chandra} images reveal
a two-component X-ray source consisting of a relatively  hard
strongly-absorbed  point source at the stellar position 
and softer less-absorbed X-ray emission extending along a 
bipolar jet several arcseconds from the star. The jet is also
seen in the optical.  Spectral extractions 
centered on the star capture emission from both components, 
resulting in a double-absorption spectrum. Some of the softer emission 
is due to the shocked jet while the harder point-like emission
at the  position of DG Tau A is undoubtedly of magnetic origin.

It is conceivable that a similar phenomenon could be responsible for 
the double-absorption X-ray spectrum detected here for IRS-1. 
It is unavoidable that our spectral extraction captures a rather
large region around the X-ray source and could sample shocked wind
or outflow emission that is offset by several arc-seconds from the 
central source. At d = 400 pc,
the {\em XMM-Newton} PSF (FWHM $\approx$ 4.$''$3) corresponds 
to FWHM $\approx$ 1700 AU and our nominal  source extraction
radius  R$_{e}$ = 15$''$ corresponds to a radius of 6000 AU (0.03 pc).
The velocity  of the  bipolar molecular outflow 
($\approx$70 km s$^{-1}$; Bally 1982) is too low to produce shock-heated
plasma at the temperatures inferred from the X-ray spectral 
fits. However, a higher velocity stellar wind or jet could 
suffice. The presence of a  compact HII region and luminosity
considerations suggest that IRS-1 may be an embedded 
$\sim$B0 - B5 star (Wa93) which could indeed have
already developed a strong wind. Also, IRS-3 lies within 
our R$_{e}$ = 15$''$ extraction region and it is known to
have a radio jet (Torreles et al. 1998).

Simple shock-heating models (Krolik \& Raymond 1985; 
Raga et al. 2002) give a predicted shock temperature
T$_{s}$ $\approx$ 1.5 $\times$ 10$^{5}$($v_{s}$/100 km s$^{-1}$)$^2$ K,
where $v_{s}$ is the shock speed relative to the upstream flow.
In order to reach X-ray temperatures comparable to that of the
cool X-ray component T$_{1}$ = 8.7 [2.3 - 11.6] MK (Table 2),
shock speeds  $v_{s}$ $\approx$ 760 [390 - 880] km s$^{-1}$ would be
required. These are fast shocks  but perhaps within reason if
IRS-1 is an embedded early B-type star. The above shock
speed is only about half the terminal wind speed of a B3 V star
and about one-third that of a B0 V star (Table 4 of 
Cassinelli et al. 1994). 

\subsection{Comments on X-rays from Wind Shocks}
X-rays from radiation-driven wind
shocks in early-type stars are predicted on theoretical grounds
(Lucy 1982; Lucy \& White 1980).
The expected X-ray temperatures are $<$1 keV so this process
could at best account only for the cooler plasma seen in the 
two-component spectrum. However, the moderately high 
X-ray absorption, plasma temperature, 
and required shock speed for the cool X-ray component noted
above stretch the limits of what the radiative wind shock model 
can accommodate.

A more interesting  possibility is that the {\em hot} plasma
is shock related. The impact of a high-velocity wind or jet-like
outflow from IRS-1 on a close companion or other obstruction
could produce high-temperature X-ray plasma in a colliding wind shock.
The colliding wind picture is usually invoked to explain high-temperature
X-ray emission  in massive close binaries such as WR $+$ O systems
(e.g. Skinner et al. 2001), but in the present case
the high-speed wind of the embedded star might actually be shocking
on dense surrounding material. Interestingly, Seth et al. (2002)
have suggested that the walls of the cavity surrounding IRS-1 might
be the interface between outflowing and infalling material.
A similar picture in which a high-velocity stellar wind is shocking
on dense surrounding clumps has been discussed by Kitamura
et al. (1990).

The  maximum colliding wind shock  temperature in the adiabatic 
case is kT$_{cw}$ $\approx$ 
1.95 ($v_{\perp,s}$/1000 km s$^{-1}$)$^2$ keV, where 
$v_{\perp,s}$ is the velocity component normal to the
shock interface (Luo, McCray, \& MacLow 1990). Thus, a
B0 V star with a terminal wind speed $v_{\infty}$ $\approx$
2500 km s$^{-1}$ could in principle produce very hot 
X-ray plasma at kT$_{cw}$ $\approx$ 12 keV, similar to the
value kT$_{2}$ $\approx$ 11 keV inferred for the hot component
in the X-ray spectrum (Table 2). The colliding wind model, if
relevant, would of course  need to account for the X-ray luminosity
(Table 2). The X-ray luminosity predicted for a colliding wind system 
is a sensitive function of mass-loss parameters and orbital separation
(Luo et al. 1990). In the absence of such  information for IRS1,
a meaningful comparison with theory cannot be made and the 
colliding wind scenario thus remains quite speculative.

\subsection{Magnetic Processes in Massive Young Stars?}
The hard X-ray continuum detected in IRS-1 extends up to 
at least 8 keV. Such  hard emission represents 
an extreme case for wind-generated shocks but is not
uncommon for X-ray emission from magnetic reconnection
processes that are observed in low-mass pre-main sequence
stars (T Tauri stars) and even some low-mass 
protostars (Imanishi, Koyama, \& Tsuboi 2001).

Some additional support for  magnetic behavior in massive
young stellar objects comes from the {\em Chandra} detection
of the high-mass embedded object Mon R2 IRS-2 (Kohno et al. 2002).
The  X-ray absorption, temperature, and X-ray luminosity reported
for Mon R2 IRS-2 are strikingly similar to the values we determine
for the hot component of NGC 2071 IRS-1. However, the fluorescent
Fe line was not seen in Mon R2 IRS-2. More importantly, the 
X-ray emission of Mon R2 IRS-2 was found to be variable on
timescales of a few 10$^{4}$ s, as was that of at least two
other embedded high-mass YSOs in Mon R2. The presence of such 
short-term variability in combination with very hot plasma 
is a strong argument in favor 
of magnetic processes.

Because of the close similarities in X-ray spectral properties
between NGC 2071 IRS-1 and Mon R2 IRS-2 (apart from the lack of
fluorescent Fe in the latter), it is quite possible that their
high-temperature X-ray emission arises from similar processes. 
If the emission
is indeed of  magnetic origin, then  the key questions are
whether the emission is due to as yet undetected late-type
companions or the massive YSOs themselves. If the emission is
intrinsic to the embedded high-mass objects, then the 
theoretical challenge will be to determine if the fields are
internally generated (and by what mechanism), or instead primordial.

\subsection{X-rays from Inverse Compton Scattering}
In conclusion, we comment briefly on the possibility
that the X-ray continuum  emission of the hard source
near IRS-1 is nonthermal. As we have noted,
the spectrum can be fitted with an absorbed power-law
continuum plus a Gaussian Fe line (model B in Table 2),
but the inferred absorption is much less than expected
toward IRS-1.
The production of nonthermal X-rays from 
OB star winds has been considered by Chen \&
White (1991). In their model, hard X-rays above 
2 keV can be produced by inverse Compton scattering of
stellar UV photons by relativistic electrons accelerated in
wind shocks near the star. 

To explore the nonthermal possibility further, we fitted the
pn X-ray spectrum with a model consisting of an absorbed
cool thermal component and an absorbed power-law component
plus a Gaussian line. The cool thermal component is intended 
to model any soft radiative wind shock emission and the
power-law component models nonthermal emission from inverse
Compton scattering. Different absorption columns were
allowed for the thermal and power-law components. This
model gives a statistically acceptable fit of the spectrum
($\chi^2$/dof = 11.3/11) with a cool thermal plasma temperature 
kT$_{1}$ $\approx$ 0.2
keV and a photon power-law index $\alpha_{ph}$ = $+$0.6.
However, the fit converges to a very large emission measure
for the cool thermal component which leads to a 
high unabsorbed luminosity L$_{\rm X}$ $\sim$ 10$^{34}$
ergs s$^{-1}$. The inferred presence of a very soft thermal
component with extremely high emission measure viewed under 
high absorption log N$_{\rm H,1}$ $\approx$ 22.9 cm$^{-2}$ 
is likely a result of fitting the data with an inappropriate 
model and the fit results seem  unphysical. We thus
do not favor such a  hybrid thermal $+$ nonthermal emission 
model based on the existing data but the model would be worth
reconsidering if higher quality spectra are obtained.

\section{Summary and Outlook}
We have presented results of the first X-ray observation
centered on the core region of the infrared cluster in 
NGC 2071. The most important (and unanticipated) result of this study
is the unusual X-ray spectrum of the source detected within  
1$''$ of the massive young stellar object IRS-1. The small 
positional offset and strong X-ray absorption inferred from 
thermal spectral models make an association of this X-ray source
with IRS-1 likely. 

The high X-ray temperature implied by
thermal models is characteristic of magnetic processes
and raises the intriguing possibility that magnetic fields
play an important role in X-ray production in young 
high-mass stars. Kohno et al. (2002) were led to a similar
conclusion based on {\em Chandra} observations of massive
young stars in Mon R2. However, the role played by any as
yet unresolved close companions in the X-ray emission
process is unknown.

A higher resolution {\em Chandra} observation of NGC 2071-IR
now pending will answer several key questions regarding the
origin of the unusual X-ray emission. Specifically, 
{\em Chandra's} arc-second resolution will place tighter
constraints on the position of the X-ray peak relative
to IRS-1 and will determine if nearby objects such as 
IRS-2, IRS-3, or IRS-5 contribute to the X-ray emission.
The longer {\em Chandra} exposure will also provide
a more definitive test for variability on short timescales,
a  key discriminant between magnetic and shock processes.

\acknowledgments

This research was supported by NASA grants NNG05GK52G and NNG05GE69G.
Work at PSI (M.G.) was supported by  Swiss National Science 
Foundation grant 20-66875.01. M.A. acknowledges support from
NASA grant NNG05GK35G and a Swiss National Science Foundation
Professorship (PP002-110504). This work is based on observations
obtained with {\em XMM-Newton}, an ESA science mission with instruments
and contributions directly funded by ESA states and the USA (NASA).
We have utilized data products from the Two Micron All-Sky Survey
(2MASS), which is a joint project of the University of Massachusetts
and IPAC/CalTech. 


\clearpage

\begin{deluxetable}{lllcclllc}
\tabletypesize{\scriptsize}
\tablewidth{0pt}
\tablecaption{X-ray Sources in NGC 2071\tablenotemark{a} }
\tablehead{
\colhead{No.}      &
\colhead{R.A.}     &
\colhead{Decl.}    &
\colhead{Rate}     &   
\colhead{$<$E$>$}   &
\colhead{Flux}  &
\colhead{Identification} & 
\colhead{ K$_{s}$ } &
\colhead{offset}               \\
\colhead{   }    &
\colhead{(J2000)     }    &
\colhead{(J2000)     }    &
\colhead{(c/s)     }   &
\colhead{(keV)  }   &
\colhead{(ergs/cm$^{2}$/s)}    &
\colhead{       }   &
\colhead{(mag)} &
\colhead{  ($''$)     }
}
\startdata
1      & 05 46 44.10     & $+$00 18 02.6   & 1.49E-02  & 1.66             & 8.18E-14 & 2MA J054644.08$+$001803 & 10.2 & 0.6 \nl
2      & 05 46 44.91     & $+$00 24 51.3   & 5.69E-04  & 2.33             & 1.03E-14\tablenotemark{e} & ... & ... & ... \nl
3      & 05 46 45.84     & $+$00 17 00.3   & 1.87E-03  & 2.34             & 1.46E-14 & ... & ... & ... \nl
4      & 05 46 51.51     & $+$00 19 20.9   & 1.34E-02  & 1.68             & 6.18E-14 & 2MA J054651.48$+$001921 & 11.8 & 0.6 \nl
5\tablenotemark{d}       & 05 46 51.80     & $+$00 19 39.4   & 1.46E-03   & 1.89 & 1.99E-14\tablenotemark{e} &  2MA J054651.85$+$001938 & 11.6 & 1.1 \nl
6\tablenotemark{c}       & 05 46 52.51     & $+$00 19 59.2   & 6.02E-04   & 1.80 & 1.64E-14\tablenotemark{e} & ... & ... & ... \nl
7      & 05 46 53.47     & $+$00 26 32.3   & 8.70E-04  & 2.62             & 4.81E-15\tablenotemark{e}        & ... & ... & ... \nl
8      & 05 46 56.56     & $+$00 20 52.5   & 1.07E-02  & 1.97             & 4.81E-14 & 2MA J054656.53$+$002052 & 10.6 & 0.5 \nl
9\tablenotemark{b}       & 05 46 58.46     & $+$00 22 35.3   & 4.98E-04   & 1.93 & 6.80E-15\tablenotemark{e} & 2MA J054658.37$+$002236 & 14.1 & 1.7 \nl
10      & 05 46 58.71    & $+$00 20 29.0   & 3.04E-03  & 2.04             & 1.02E-14 & Wa93-51; 2MA J054658.59$+$002029 & 12.2 & 1.8 \nl
11     & 05 46 59.03     & $+$00 24 57.8   & 1.44E-03  & 3.27             & 1.74E-14 & 2MA J054659.03$+$002457 & 12.4 & 0.1 \nl
12     & 05 47 03.38     & $+$00 23 23.5   & 1.98E-02  & 2.29             & 1.39E-13 & 2MA J054703.31$+$002323 & 10.4 & 1.0 \nl
13     & 05 47 04.78     & $+$00 21 43.5   & 8.77E-03  & 4.62             & 1.29E-13 & IRS-1; 2MA J054704.78$+$002142 & 11.2 & 0.7 \nl
14     & 05 47 05.00     & $+$00 18 32.0   & 1.76E-02  & 1.88             & 8.92E-14 & 2MA J054704.94$+$001831 & 10.4 & 0.9 \nl
15\tablenotemark{b}      & 05 47 05.21     & $+$00 23 08.4   & 4.02E-03   & 2.64 & 7.01E-14 & ... & ...  & ... \nl
16\tablenotemark{h}      & 05 47 05.27     & $+$00 22 53.7   & 2.41E-02  & 3.19             & 1.94E-13 & 2MA J054705.25$+$002253 & 12.4 & 0.3 \nl
17\tablenotemark{b}      & 05 47 05.56     & $+$00 22 11.4   & 9.30E-04   & 2.87 & 1.27E-14\tablenotemark{e} & ...& ...  & ... \nl
18     & 05 47 06.26     & $+$00 24 53.5   & 2.30E-02    & 2.51           & 1.74E-13 & 2MA J054706.26$+$002454 & 10.7 & 0.5 \nl
19     & 05 47 07.29     & $+$00 19 32.0   & 5.79E-02    & 2.70           & 4.27E-13 & Lk308, 2MA J054707.26$+$001932 & 8.3 & 0.4 \nl
20     & 05 47 07.66     & $+$00 17 40.9   & 1.11E-01    & 2.01           & 5.79E-13 & 2MA J054707.64+001740 & 8.8 & 0.3 \nl
21\tablenotemark{d}      & 05 47 07.93     & $+$00 17 55.5   & 1.30E-02   & 1.74 & 1.60E-13 & 2MA J054707.91$+$001756\tablenotemark{f}& 6.3 & 0.5 \nl
22     & 05 47 08.92     & $+$00 20 26.5   & 2.17E-03    & 2.50           & 1.20E-14\tablenotemark{e} & ... & ... & ... \nl
23     & 05 47 09.77     & $+$00 22 35.5   & 5.55E-03    & 2.90           & 4.71E-14 & 2MA J054709.77$+$002236 & 12.0 & 0.5 \nl
24     & 05 47 10.21     & $+$00 20 08.9   & 1.74E-03    & 3.11           & 9.62E-15\tablenotemark{e} & ...\tablenotemark{g}     & ...   & ... \nl
25\tablenotemark{c}      & 05 47 10.33     & $+$00 17 22.2   & 1.73E-03   & 2.29 & 2.36E-14\tablenotemark{e} & 2MA J054710.36$+$001721 & 12.0 & 0.7 \nl
26     & 05 47 10.59     & $+$00 18 44.3   & 2.11E-03    & 2.16           & 8.37E-15 & USNO J054710.63$+$001845 & ... & 1.3 \nl
27     & 05 47 10.70     & $+$00 26 19.9   & 5.02E-04    & 3.24           & 2.78E-15\tablenotemark{e} & ... & ... & ... \nl
28     & 05 47 10.91     & $+$00 19 14.9   & 3.31E-03    & 2.13           & 1.82E-14 & Lk310, 2MA J054710.98$+$001914 & 9.1 & 1.1 \nl
29     & 05 47 12.03     & $+$00 17 57.6   & 8.87E-03    & 1.97           & 6.93E-14 & 2MA J054712.02$+$001756 & 10.6 & 0.8 \nl
30\tablenotemark{c}      & 05 47 12.95     & $+$00 22 06.4   & 2.59E-03   & 2.40 & 3.41E-14 & 2MA J054712.92$+$002206 & 10.5 & 0.5 \nl
31\tablenotemark{b}      & 05 47 15.51     & $+$00 18 43.2   & 6.64E-04   & 3.40 & 9.06E-15\tablenotemark{e} & ... & ... & ... \nl
32\tablenotemark{c}      & 05 47 15.86     & $+$00 23 21.4   & 9.63E-04   & 3.26 & 1.31E-14\tablenotemark{e} & 2MA J054715.92$+$002320 & 13.3 & 1.4 \nl
33     & 05 47 19.27     & $+$00 19 19.5   & 1.91E-03    & 2.26           & 6.04E-15 & 2MA J054719.19$+$001920 & 11.7 & 1.7 \nl

\enddata

\tablenotetext{a}{
Notes:~Data are from the {\em XMM-Newton} observation on 30-31 March 2005
and include sources within a $\approx$10$'$ $\times$ 10$'$ region centered
on NGC 2071-IRS1.
All quantities are computed using events in the 0.5 - 7.5 keV range.
Usable exposure times are 29,886 s for pn and 30,114 s per MOS.
Data are from EPIC pn  unless otherwise noted. The count rate is based
on events extracted within a circular region of radius 15$''$ (68\% EEF)
centered on the source and is background subtracted. A smaller source
region of radius $\approx$10$''$ was used for sources 5,16, and 21 
due to source crowding. The absorbed flux 
(0.5 - 7.5 keV)  is based on spectral fits using an absorbed 
solar-abundance 1T or 2T optically thin plasma model ($apec$) in XSPEC,
except as noted for faint sources.  $<$E$>$ is the mean photon energy.
Candidate identifications
lie within 2$''$ of the X-ray position and are based (in order of
preference) on searches of the  2MASS (2MA), HST GSC v2.2, and USNO B1 
electronic data bases. The 2MASS K$_{s}$ magnitude is given 
for sources with 2MASS identifications.
The quoted {\em offset} is the positional offset
between the X-ray and 2MASS or USNO positions. } 

\tablenotetext{b}{Based on  MOS2 events. Source not detected in pn.} 

\tablenotetext{c}{Based on  MOS2 events. Source lies near pn CCD gap.}

\tablenotetext{d}{Based on  MOS2 events. Second source lies nearby.}

\tablenotetext{e}{Absorbed flux (0.5 - 7.5 keV) is from PIMMS based on 
the quoted count rate and
an assumed Raymond-Smith model with N$_{\rm H}$ = 10$^{22}$ cm$^{-2}$ 
and plasma temperature kT = 2 keV.}

\tablenotetext{f}{The star HDE 290861 is offset 0.$''$8 from the 
                  X-ray position.}

\tablenotetext{g}{IR source 29 in Walther et al. 1993 is offset by 2.$''$4 from
  the X-ray position.}

\tablenotetext{h}{MOS images show a faint source located $\approx$17$''$ north of 
                  source 16.}

\end{deluxetable}


\clearpage

\begin{deluxetable}{lll}
\tabletypesize{\scriptsize}
\tablewidth{0pc}
\tablecaption{{\em XMM-Newton} Spectral Fits for IRS-1 
   \label{tbl-1}}
\tablehead{
\colhead{Parameter}      &
\colhead{ }        &
\colhead{  }
}
\startdata
Model\tablenotemark{a}                &           A         &         B                        \nl
Emission                              & thermal $+$ line    & power-law $+$ line               \nl
N$_{\rm H,1}$ (10$^{22}$ cm$^{-2}$)   & 3.8  [2.2 - 10.]    & 1.0 [0.19 - 2.3]               \nl
kT$_{1}$ (keV)                        & 0.75 [0.20 - 1.0]   & ...               \nl
norm$_{1}$ (10$^{-4}$)\tablenotemark{b} & 2.1 [0.4 - ...]   & 0.04 [0.02 - 0.06]           \nl 
N$_{\rm H,2}$ (10$^{23}$ cm$^{-2}$)   & 1.6 [0.48 - 3.0]    & ...               \nl
kT$_{2}$ (keV)                        & 10.8 [5.8  - ...]   & ...               \nl
norm$_{2}$  (10$^{-4}$)\tablenotemark{b} & 1.4 [0.76 - 2.3] & ...          \nl
$\alpha_{ph}$                          & ...                 & 0.55 [0.26 - 0.85]             \nl
E$_{line}$ (keV)                      & 6.43 [6.37 - 6.53]  & 6.48 [6.42 - 6.54]          \nl
$\sigma_{line}$ (keV)                 & 0.14 [0.06 - 0.26]  & 0.21 [0.14 - 0.28]                   \nl
norm$_{line}$ (10$^{-6}$)             & 4.5 [2.9 - 6.5]     & 5.5 [4.2 - 6.9]                          \nl     
$\chi^2$/dof                          & 10.7/11             & 14.3/14                                            \nl
$\chi^2_{red}$                        & 0.97                & 1.02                                              \nl
F$_{\rm X}$ (10$^{-13}$ ergs cm$^{-2}$ s$^{-1}$)          & 1.28 (7.93) & 1.27 (1.39)                          \nl
F$_{\rm X,1}$ (10$^{-13}$ ergs cm$^{-2}$ s$^{-1}$)        & 0.13 (4.99) & ...                            \nl
F$_{\rm X,line}$ (10$^{-14}$ ergs cm$^{-2}$ s$^{-1}$)     & 3.86 (4.03) & 3.56 (3.56)                          \nl
L$_{\rm X}$ (10$^{31}$  ergs s$^{-1}$)                    & 1.5         & 0.27                                \nl
L$_{\rm X,1}$ (10$^{31}$  ergs s$^{-1}$)                  & 0.95         & ...                                \nl
\enddata
\tablecomments{
Based on  XSPEC (vers. 12.3.0) fits of the background-subtracted EPIC pn spectrum binned 
to a minimum of 15 counts per bin using 29.9 ksec of low-background exposure. Thermal
emission was modeled with a solar abundance (Anders \& Grevesse 1989) 
$apec$ optically thin plasma model in XSPEC.  
The tabulated parameters
are absorption column density (N$_{\rm H}$), plasma energy (kT),
component normalization (norm), photon power-law index ($\alpha_{ph}$), Gaussian line
centroid energy (E$_{line}$), and line width ($\sigma_{line}$ = FWHM/2.35).
Solar abundances are referenced to  Anders \& Grevesse (1989).
Square brackets enclose 90\% confidence intervals and an ellipsis means that 
the algorithm used to compute confidence intervals did not converge.
The total X-ray flux (F$_{\rm X}$) and flux of the low-absorption component
(F$_{\rm X,1}$) are the absorbed values in the 0.5 - 7.5 keV range, followed in 
parentheses by  unabsorbed values. The continuum-subtracted fluorescent Fe  line flux
(F$_{\rm X,line}$) is measured in the 6.2 - 6.6 keV range.
The unabsorbed total luminosity L$_{\rm X}$ (0.5 - 7.5 keV)  and cool-component
luminosity L$_{\rm X,1}$ (0.5 - 7.5 keV) assume a  
distance of 400 pc. } 
\tablenotetext{a}{Model A:~N$_{\rm H,1}$$\cdot$kT$_{1}$ $+$ 
                           N$_{\rm H,2}$$\cdot$kT$_{2}$ $+$ GAUS;
                  Model B:~N$_{\rm H,1}$$\cdot$PL $+$ GAUSS}
\tablenotetext{b}{For thermal $vapec$ models, the norm is related to the emission measure (EM) by
                  EM = 4$\pi$10$^{14}$d$_{cm}^2$$\times$norm, where d$_{cm}$ is the stellar
                  distance in cm.}
\end{deluxetable}


\clearpage

\begin{figure}
\figurenum{1}
\epsscale{1.0}
\includegraphics*[width=9.5cm,angle=-90]{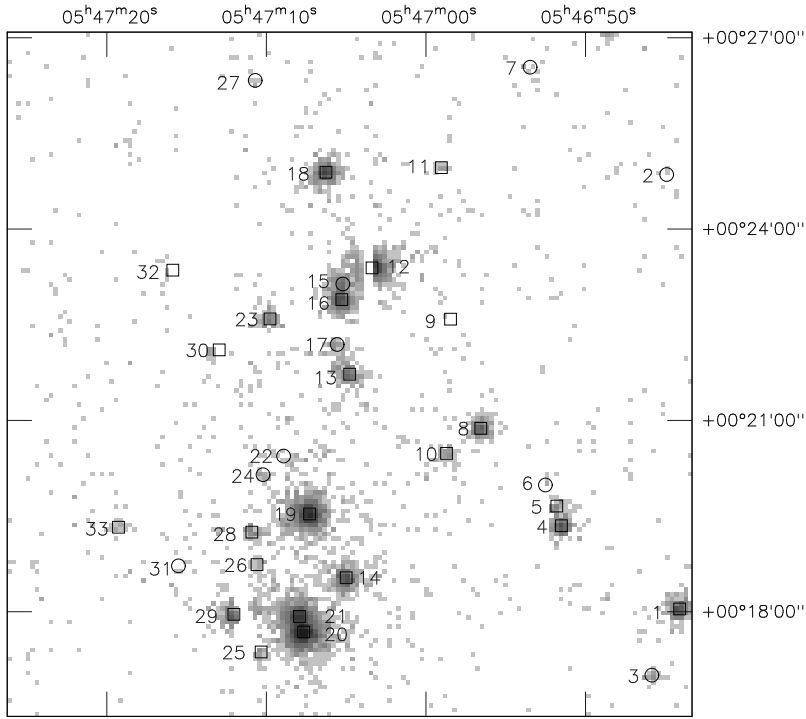}
\caption{Broad-band {\em XMM-Newton} EPIC pn X-ray image of the $\approx$10$'$
$\times$ 10$'$ central region in NGC 2071 (0.5 - 7 keV; 29.9 ksec 
usable exposure; rebinned to a  pixel size of 4.$''$4; log scale; 
J2000.0 coordinates). Numbered X-ray sources correspond to Table 1.
Boxes enclose X-ray sources with identified counterparts and circled
sources lack counterparts. }
\end{figure}

\clearpage

\begin{figure}
\figurenum{2}
\epsscale{1.0}
\includegraphics*[width=9.5cm,angle=-90]{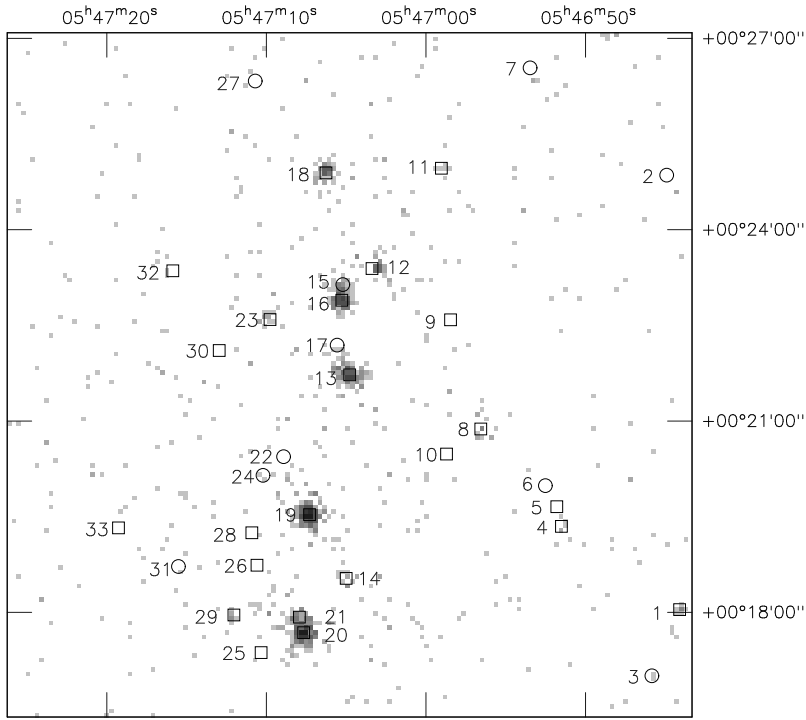}
\caption{Same as Figure 1, but restricted to the hard
4 - 7 keV band. }
\end{figure}

\clearpage

\begin{figure}
\figurenum{3}
\epsscale{1.0}
\includegraphics*[width=9.5cm,angle=-90]{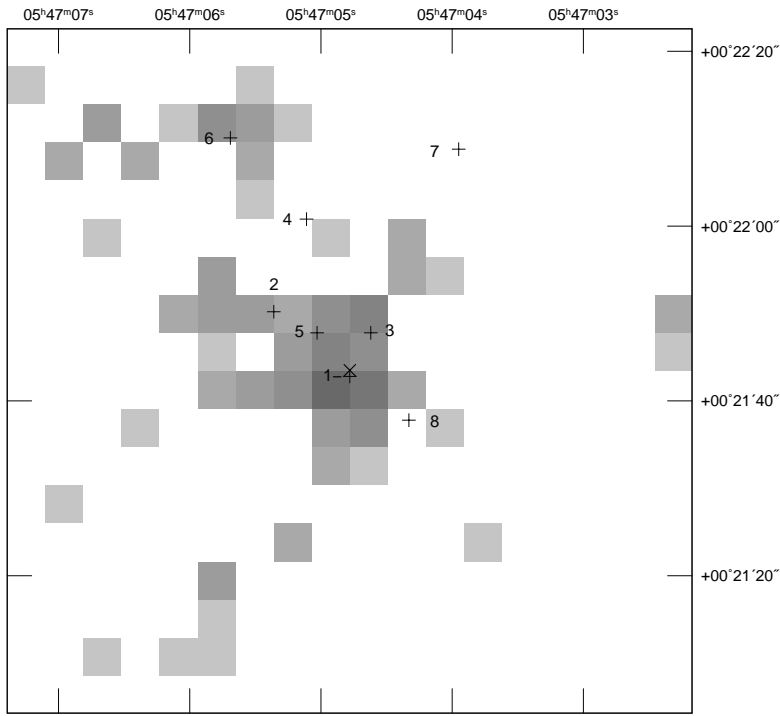}
\caption{Zoomed EPIC MOS1$+$2 X-ray image of the central 
NGC 2071-IR subgroup (0.5 - 7 keV; rebinned to a pixel size of 4.$''$4; 
log scale; J2000.0). 
Crosses  show IR positions of IRS 1-8 (Wa93) and $\times$ is
the centroid of the X-ray source associated with IRS-1.}
\end{figure}

\clearpage

\begin{figure}
\figurenum{4}
\epsscale{1.0}
\includegraphics*[width=9.5cm,angle=-90]{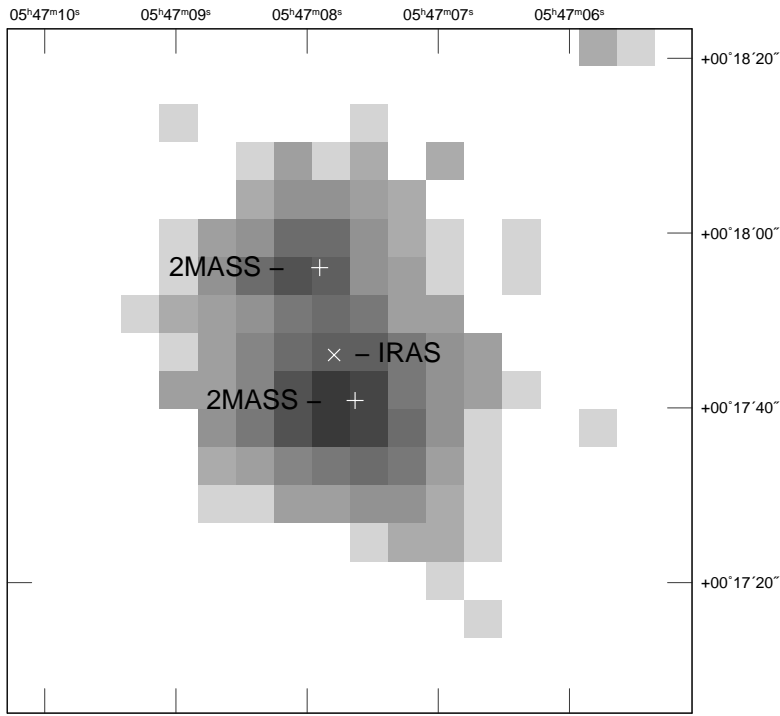}
\caption{EPIC MOS1$+2$ image (0.5 - 7 keV) of the 
 close pair sources 20 (south) and 21 (north). 
 Positions of 2MASS sources
 are marked with crosses ($+$). Source 21 is 
 possibly associated with the B star HDE 290861 
 (or its close companion) which lies at an offset of 0.$''$8 from the 
 X-ray peak.   The position of IRAS
 05445$+$0016 ($\times$) has a 95\% uncertainty ellipse of 
 47$''$ $\times$ 7$''$ (semi-major $\times$ semi-minor axes)
 with the ellipse major axis at PA = 89$^{\circ}$. }
\end{figure}

\clearpage

\begin{figure}
\figurenum{5}
\epsscale{1.0}
\includegraphics*[width=9.5cm,angle=-90]{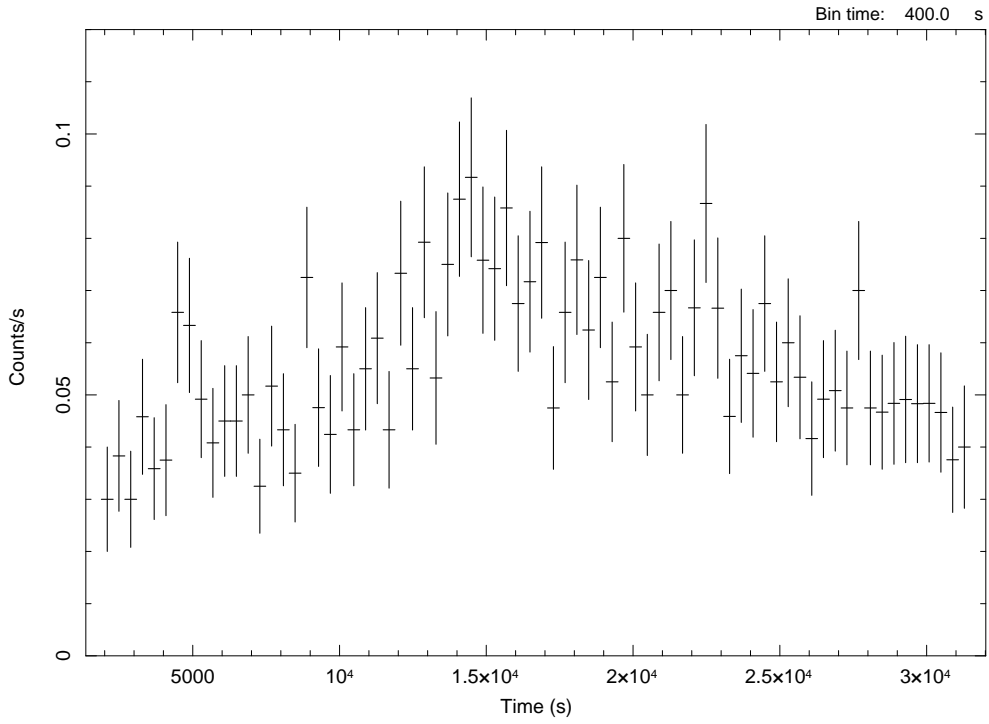}
\caption{Background-subtracted EPIC pn light curve of
the emission-line star LkH$\alpha$ 308 (source 19 = 
XMM J054707.29$+$001932) in the   
0.5 - 7 keV range. The bin size is 400 s.}
\end{figure}

\clearpage

\begin{figure}
\figurenum{6}
\epsscale{1.0}
\includegraphics*[width=9.5cm,angle=-90]{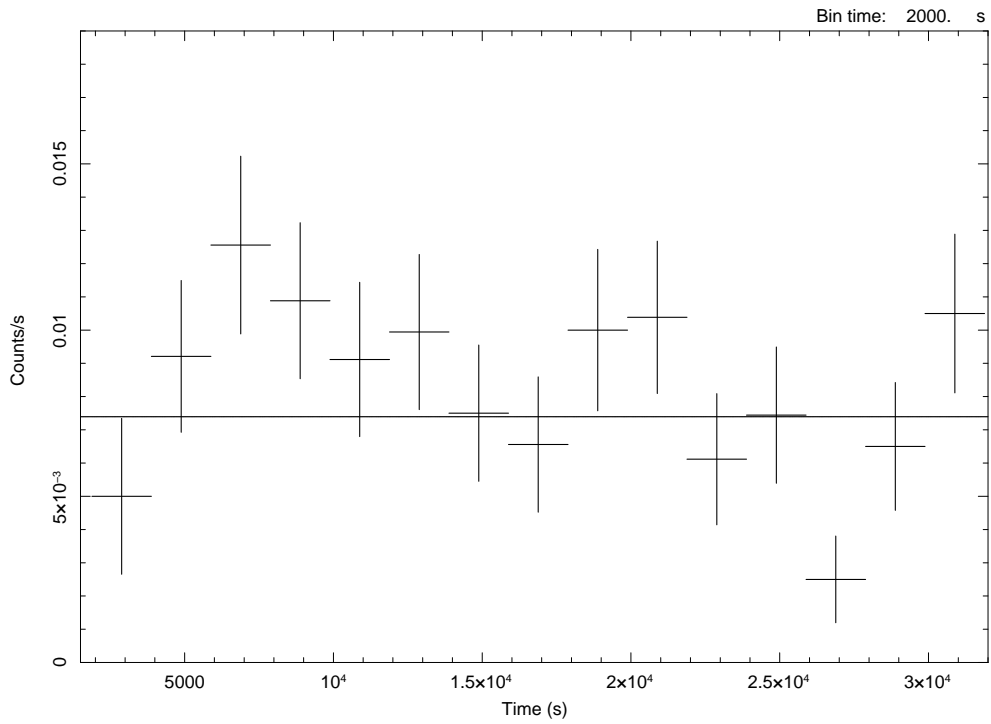}
\caption{Background-subtracted EPIC pn light curve of 
XMM J054704.78$+$002143 (offset by 0.$''$7 from IRS-1)
in the 0.5 - 7 keV range. The bin size is 2000 s and
the solid line is a best-fit constant count rate model.}
\end{figure}

\clearpage

\begin{figure}
\figurenum{7}
\epsscale{1.0}
\includegraphics*[width=9.5cm,angle=-90]{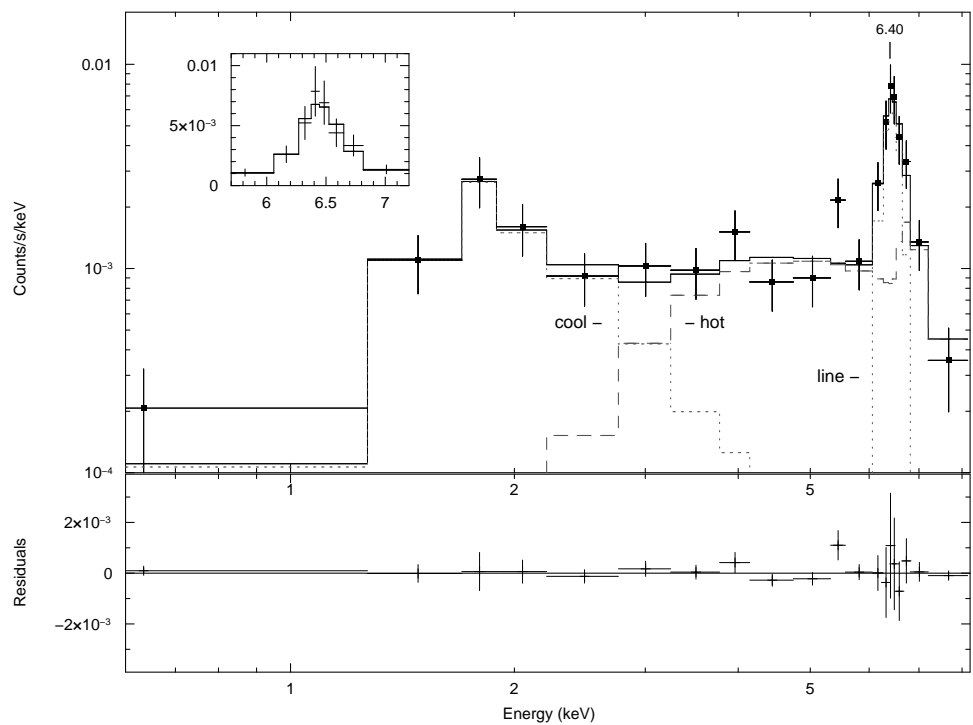}
\caption{Background-subtracted EPIC pn spectrum of 
XMM J054704.78$+$002143 (offset by 0.$''$7 from IRS-1) 
binned to a minimum of 
15 counts per bin (262 net counts in the 0.5 - 7.5 keV
range; extraction radius  R$_{e}$ = 15$''$).
The spectrum is dominated
by a strong fluorescent Fe  emission line near 6.4 keV. 
The solid line is a best-fit double-absorber thermal model
(model A in Table 2) with a cool moderately absorbed 
component (dotted line) and a hot heavily absorbed 
component (dashed line), plus a Gaussian line (dotted). 
The inset shows the Fe line on a linear axis scale. } 
\end{figure}

\clearpage


\begin{thebibliography}{}
\bibitem[]{} Anders, E. \& Grevesse, N. 1989, Geochim. Cosmochim. Acta,
             53, 197
\bibitem[]{} Anthony-Twarog, B. 1982, \aj, 87, 1213
\bibitem[]{} Aspin, C., Sandell, G., \& Walther, D.M. 1992. \mnras,
             258, 684
\bibitem[]{} Bally, J. 1982, \apj, 261, 558
\bibitem[]{} Brown, A.G.A., de Geus, E.J., \& de Zeeuw, P.T. 1994, 
             \aap, 289, 101
\bibitem[]{} Butner, H.M., Evans, N.J., Harvey, P.M., Mundy, L,G.,
             Natta, A., \& Randich, M.S. 1990, \apj, 364, 164
\bibitem[]{} Cassinelli, J.P., Cohen, D.H., MacFarlane, J.J., Sanders, W.T.,
             \& Welsh, B.Y. 1994, \apj, 421, 705
\bibitem[]{} Chen, W. \& White, R.L. 1991, \apj, 366, 512
\bibitem[]{} Clark, F.O., 1991 \apjs, 75, 611
\bibitem[]{} Gorenstein, P., 1975 \apj, 198, 95
\bibitem[]{} G\"{u}del, M., Skinner, S.L., Briggs, K.R., Audard, M.,
             Arzner, K., \& Telleschi, A. 2005, \apj, 626, L53
\bibitem[]{} Harju, J. et al., 1998 \aaps, 132, 211
\bibitem[]{} Herbig, G.H. \& Kuhi, L.V. 1963, \apj, 137, 398
\bibitem[]{} Imanishi, K., Koyama, K., \& Tsuboi, Y. 2001, \apj, 557, 747
\bibitem[]{} Kallman, T.R., 1995, \apj, 455, 603
\bibitem[]{} Kitamura, Y., Kawabe, R., Yamashita, T., \& Hayashi, M.,
             1990 \apj, 363, 180
\bibitem[]{} Kohno, M., Koyama, K., \& Hamaguchi, K. 2002, \apj, 567, 423
\bibitem[]{} Krolik, J.H. \& Raymond, J.C. 1985, \apj, 298, 660
\bibitem[]{} Lada, E.A., DePoy, D.L., Evans, N.J., \& Gatley, I.,
             1991 \apj, 371, 171
\bibitem[]{} Lucy, L.B. 1982, \apj, 255, 286
\bibitem[]{} Lucy, L.B. \& White, R.L. 1980, \apj, 241, 300
\bibitem[]{} Luo, D., McCray, R., \& MacLow, M.-M. 1990, \apj,
             362, 267
\bibitem[]{} Prieur, J.-L., Oblak, E., Lampens, P., Kurpinska-Winiarska, M.,
             Aristidi, E., Koechlin, L., \& Ruymaekers, G. 2001,
             \aap, 367, 865
\bibitem[]{} Raga, A.C., Noriega-Crespo, A., \& Vel\'{a}zquez, P.F.
             2002, \apj, 576, L149
\bibitem[]{} Seth, A.C., Greenhill, L.J., \& Holder, B.P. 2002, \apj, 581, 325
\bibitem[]{} Skinner, S.L., G\"{u}del, M., Schmutz, W., \& Stevens, I.R.
             2001, \apj, 558, L113
\bibitem[]{} Snell, R.L. \& Bally, J. 1986, \apj, 303, 683
\bibitem[]{} Strom, K.M., Strom, S.E., \& Vrba, F.J. 1976, \aj, 81, 308
\bibitem[]{} Str\"{u}der, L. et al. 2001, \aap, 365, L18
\bibitem[]{} Takagi, S.-I., Murakami, H., \& Koyama, K., 2002, \apj, 573, 275
\bibitem[]{} Torrelles, J.M., G\'{o}mez, J.F., Rodr\'{i}guez, L.F.,
             Curiel, S., Anglada, G., \& Ho, P.T.P. 1998, \apj, 505, 756
\bibitem[]{} Tsujimoto, M. et al. 2005, \apjs, 160, 503 
\bibitem[]{} Turner, M.J.L. et al. 2001, \aap, 365, L27
\bibitem[]{} Walther, D.M., Robson, E.I., Aspin, C., \& Dent, W.R.F. 
             1993, \apj, 418, 310 (Wa93)
\bibitem[]{} Wiramihardja, S.D., Kogure, T., Yoshida, S., Ogura, K., 
             \& Nakano, M. 1989, \pasj, 41, 155
\bibitem[]{} Zhao, B. et al. 1999, \aj, 118, 1347
\end{thebibliography}
\end{document}